\documentclass[aps,prb,,twocolumn,superscriptaddress,amsmath,showpacs,tightenlines]{revtex4}
\usepackage{epsfig}
\usepackage{txfonts}

\begin{document}

\title{High-Speed Magnetoresistive Random-Access Memory Random Number Generator Using Error-Correcting Code}

\author{Tetsufumi~Tanamoto, Naoharu~Shimomura, Sumio~Ikegawa, Mari~Matsumoto, Shinobu Fujita, and~Hiroaki~Yoda}
\affiliation{Corporate R \& D center, Toshiba Corporation,
Saiwai-ku, Kawasaki 212-8582, Japan}

\date{\today}
\begin{abstract}
A high-speed random number generator (RNG) circuit based on magnetoresistive 
random-access memory (MRAM) using an error-correcting code (ECC) post processing 
circuit is presented. ECC post processing increases the quality of randomness 
by increasing the entropy of random number. { 
We experimentally show that a small {\it error-correcting capability} circuit is sufficient for this post processing. 
It is shown that the ECC post processing circuit powerfully improves the quality of 
randomness with minimum overhead, ending up with high-speed random number 
generation. We also show that coupling with a linear feedback shift resistor is
 effective for improving randomness}.
 \end{abstract}

\maketitle
\section{Introduction}

In light of the advent of ubiquitous networks, security for mobile applications
has become more important and advanced security systems 
are required. 
For many security systems, random numbers are widely used 
to create IDs, passwords, and so on. 
An on-chip random number generator (RNG) is one of the key elements for a secure system-on-chip.  
Unpredictability is one of the most important characteristics of 
random numbers. RNGs capable of generating natural random numbers 
by means of physical 
phenomena have recently been investigated and have begun to 
replace pseudo-random numbers generated by software.
Various kinds of on-chip RNG
using physical noise signals such as random telegraph noise in transistors~\cite{Matsumoto,Brederlow,Tokunaga,Fujita,Matsu2,Uchida}, 
{noise in diode junctions, or magnetic tunnel junctions~(MTJs)}for magnetoresistive random-access memory (MRAM)~\cite{Fukushima,Yuasa,Ohno} 
have been proposed.

Because RNGs based on physical phenomena are 
affected to a greater or lesser degree by their physical environments and 
manufacturing processes, for example by variation of device size, after AD-converting physical 
noise signals, post processing using digital circuits is always needed to balance `0' and `1'  to eliminate 
periodicity and correlation with the original signals. 
The ``rejection method'' is most frequently used for {the post processing circuit} (PPC), where two serial bit 
sequences `01' and `10' are replaced by `0' and `1', whereas `00' and `11' are discarded. 
Although this method is simple and clear, when the original physical signals fluctuate, 
the balance between  `0' and `1' easily collapses and leads to a low generation rate of random numbers 
after wasting a lot of bits to create balanced random numbers. 
One of the more sophisticated methods for PPC is to apply a data compression method 
in order to increase the entropy of randomness. Lacharme~\cite{Lacharme} proposed a PPC 
using the Bose-Chaudhuri-Hocquenghem (BCH) {error-correcting code} (ECC).
He proved that derivation of probability of `0' and `1' from $1/2$, $e/2$,  can be reduced 
to $e^d/2$ with a minimal distance $d$ of the ECC code. 
However, it is unclear whether this PPC improves the quality of randomness 
sufficiently to pass commercial-level statistical tests~\cite{NIST,AIS31}. 
Moreover, compared with the noise sources, the area of an ECC circuit is generally 
larger and consumes more power. Consequently, 
the additional ECC circuit for RNGs is inclined to 
dominate the area and power of the RNG. 

\begin{figure}[!t]
\centering
\includegraphics[width=6cm]{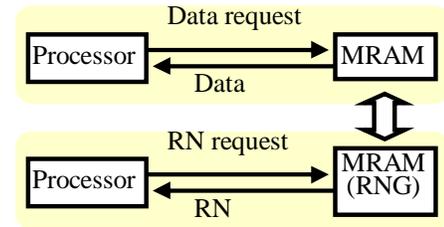}
\vspace*{3cm}
\caption{MRAM including ECC circuit is used to generate random numbers with post processing ECC,
 when processor requests random numbers. The ECC circuit for correcting memory bit error is 
shared with the post processing circuit.}
\label{processor}
\end{figure}

\begin{figure}[b]
\centering
\includegraphics[width=5cm]{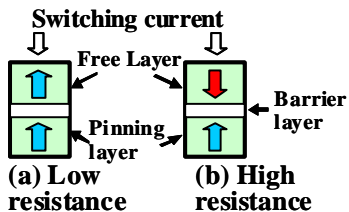}
\vspace*{3cm}
\caption{MRAM device based on perpendicular magnetic tunnel junction (MTJ). 
A random number is generated at a  switching current where the magnetization of 
the free layer changes at 50 \%.  To reset the free layer, 800~mV is applied to MTJ.}
\label{fig_mram}
\end{figure}

In this paper, we present a novel MRAM-based RNG, where, in addition to the conventional memory function 
of MRAM, the memory system is also used as 
the physical noise signal, and ECC circuits used for correcting MRAM memory bit error are also converted to 
PPC when a random number is required (Fig.~\ref{processor}). MRAM produces a random number by reducing 
the programming current to cause writing errors intentionally in the RNG mode (Fig.~\ref{fig_mram}). 
The sharing of the memory system including ECC circuits with RNG leads to a 
reduction of circuit area and an efficient secure memory system for mobile applications.

The remainder of this paper is organized as follows: in \S 2, we describe
our general idea of PPC using ECC circuit including 
an explanation of the conventional ECC procedure. 
In \S 3, we present our MRAM system to create RNG.
In \S 4. we show our experimental results and analyze 
the ECC PPC process. \S 5 is a discussion part. 
Our conclusions are summarized in \S 6.

\section{Post Processing Using ECC Circuit}
\subsection{RNG PPC circuit}
The data compression by PPC using ECC increases the entropy of random numbers. 
A block of $n$ bits is transferred into $k$ bits by this circuit~($n>k$). 
Figure \ref{processor} shows a schematic of a proposed post processing circuit 
shared by the conventional error-correction linear block. 
The ECC circuit is usually used for the error correction of memory bits.
The key point of our system is that we can change the role of a conventional ECC circuit to PPC
by adding only a small number of transistors (Figs.~3 and 4).
First, let us remember the conventional ECC.
A given message data stream is broken up into blocks, such as $k$-symbol data block ${\bf m}=(m_0, m_1, ..., m_{k-1})$.
A generating matrix ${\bf G}$ encodes a given message ${\bf m}$ to a code 
${\bf c}={\bf m}{\bf G}$.  In a simple form, ${\bf G}$ is the $n \times k$ generating matrix 
constituted from a generator polynomial  as 
{
\begin{equation}
G(x)=g_0+g_1x+...+g_{n-k}x^{n-k}, 
\end{equation} 
}
such as 
\begin{equation}
{\bf G}=\left(
\begin{array}{ccccccc}
g_{n-k} &  g_{n-k-1} &  \cdots     & g_0      & 0    & \cdots & 0          \\
    0    &  g_{n-k}    &  g_{n-k-1} &  \cdots & g_0  &          &   \vdots  \\
\vdots &              &               &            &       &   \ddots     &  0  \\
    0    &  \cdots   &        0     &     g_{n-k} & g_{n-k-1}  &  \cdots   &   g_0   \\
\end{array}
\right).
\end{equation}
In the present case, the original raw random bit sequence ${\bf y}=(y_0,...,y_n)$ 
is converted into post-processed data ${\bf z}=(z_0,...,z_k)$  as
${\bf z}={\bf G}{\bf y}$. The coefficients of the generator polynomials $G(x)$ 
for binary BCH codes for $n \le 127$ are given in Table I.

\begin{figure}
\centering
\includegraphics[width=8cm]{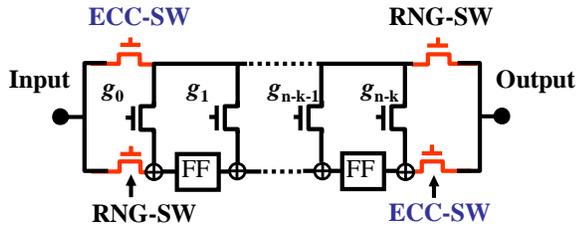}
\vspace*{3cm}
\caption{Proposed RNG post processing unit which is shared with memory bit 
error-correcting block. FF is a flip-flop element. In ECC mode, ``ECC-SW'' is ON and in RNG mode, 
``RNG-SW'' is ON. $g_0$,...,$g_{n-k}$ is determined by $G(x)$. 
For example, when $G(x)= x^5+x^2+1$,  $g_0$, $g_2$, and $g_5$ are 1 when transistors are ON states.
}
\label{PPCcircuits}
\end{figure}

\begin{figure}
\centering
\includegraphics[width=8.5cm]{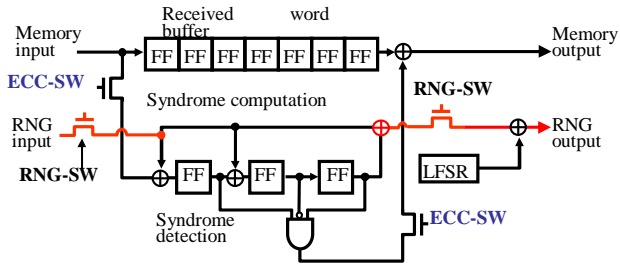}
\vspace*{4cm}\caption{Example of a proposed RNG post processing unit shared with 
ECC decoder. ECC for RNG is easily changed by switching RNG-SW 
and ECC-SW. Example for (7,4) code. We can also apply the RNG post processing unit 
to the ECC encoder block.}
\label{decode}
\end{figure}

\begin{table}[h]
\caption{BCH codes we used. The numbers of $G(x)$ show coefficients of generator polynomials 
\cite{Wicker}. For example, 45 $\rightarrow$100,101$\rightarrow$ $G(x)=x^5+x^2+1$. }
\label{BCHcode}
\begin{tabular}{cccl}
\hline
$n$   &  $k$ & $t$ & $G(x)$\\
\hline\hline
31 & 26 & 1 & 45 \\
31 & 21 & 2 & 3551 \\
31 & 16 & 3 & 107657 \\
31 &  5  & 5 & 5423325 \\
\hline
63 & 57 & 1 & 103 \\
63 & 51 & 2 & 12471 \\
63 & 45 & 3 & 1701317 \\
63 & 39 & 4 & 166623567 \\
\hline
127 & 120 & 1 & 211 \\
127 & 113 & 2 & 141567 \\
127 & 106 & 3 & 11554743 \\
127 & 99  & 4 & 3447023271 \\
\hline
\end{tabular}
\end{table}
These encoding and decoding processes of 
cyclic linear codes are efficiently implemented using exclusive-OR gates, 
switches, and shift registers that include flip-flops (FF) connected in series.
Note that data compression
${\bf z}={\bf G}{\bf y}$ for random bits ${\bf y}$ differs from 
${\bf c}={\bf m}{\bf G}$ for code ${\bf c}$ in terms of the position of FFs.
Thus, we implement both circuits by adding switching transistors, as shown in Fig~\ref{PPCcircuits}.
Then, we can change the role of the circuit between error correction of memorized bits and that of RNG 
post processing by switching transistors expressed by ECC-SW and RNG-SW in Fig.~\ref{PPCcircuits}. 
Note also that BCH codes can be chosen appropriately by controlling  ON/OFF of transistors $g_0$,.., $g_{n-k}$.
Figure~\ref{decode} shows an implemented 
form of the shared decoder block in Fig.~\ref{PPCcircuits}~[a generator polynomial of  the (7,4) code
 is given by $G(x)=x^3+x+1$]. 
In a rejection method, the number of bits is reduced to about one-fourth of the original data 
[compression rate is about less than 1/4]. On the other hand, PPC using the $(n,k,t)$ BCH code transform $n$ bits to $k$ bits, 
and thus the compression rate is $k/n$ [$t$ is an {\it error-correcting capability}].

\subsection{Coupling with linear feedback shift resistor~(LFSR)}
Because physical RNGs originate from physical phenomena, there are some cases 
where physical RNGs does not produce desirable random numbers, such as 
electrical breakdown of stacked thin films. 
Thus, in general, some supplementary circuits are always required for commercial use\cite{AIS31}.
A linear feedback shift resistor (LFSR) is most frequently used for the support of physical RNGs. 
{LFSR, which produces a sequence of bits, is composed 
of a shift resistor and a feedback function by XOR of certain bits in the resistor.}
$N$-bit LFSR, which is constituted by $N$ FFs, 
can hold $2^N-1$ bit-long pseudo-random sequences for appropriate input data\cite{Schneier}. 
Because each FF 
includes more than 20 transistors, a smaller $N$-bit LFSR is desirable for 
reducing circuit area. Here, we investigate $N$-bit LFSR with $N \le 8$.
Concretely, we try $(1,0)$, $(2,1,0)$, $(3,1,0)$, $(4,1,0)$, $(7,1,0)$, and $(7,3,0)$ LFSRs,
{
where listing $(1,0)$ means that a new bit is generated by XORing a shifted bit with the next bit, and listing $(7,1,0)$ 
means that a seven times shifted bit and a shifted bit are XORed with the next bit, and so on.}
Figure~\ref{LFSR} shows $(2,1,0)$-LFSR and $(3,1,0)$-LFSR.
We found that small LFSR is suitable for MRAM RNG, as shown below.

\begin{table}
\caption{Typical random numbers examined here. Data $A$ is an average sample. Data $B$ is a low-quality sample. 
Data $C$ is a higher-speed sample.}
\begin{tabular}{ccc}
\hline
Data     & Balance & $t_{\rm write}$ (ns) \\
\hline\hline
$A$ & 51.1 \% `1' & 30 \\
\hline
$B$ & 27.6 \% `1' & 30 \\
\hline
$C$ & 36.3 \% `1' & 10 \\
\hline
\end{tabular}
\end{table}

\section{MRAM RNG}
MRAM cell arrays are integrated by the 130~nm front process and 240~nm back-end process \cite{Kishi,Aikawa,Yoda,Ikegawa} (Fig.~\ref{SEM}). 
An MRAM random number is generated by controlling the switching current 
such that the probability of change of magnetism of the free layer is 50 \% (Fig. 1)~\cite{Fukushima}. 
Figure \ref{pulse} shows the MRAM switching probability $P_{\rm write}$ as a function of switching current at various pulse widths\cite{Yoda}. 
In the present MRAM, the slopes of the change in switching probabilities look similar 
regardless of the pulse duration rate. This leads to the stable operation of MRAM cells. 
A closer look at this figure, in particular at the tail parts of the slopes, reveals
that the slope of $P_{\rm write}$ becomes slightly smaller for a shorter pulse width $t_{\rm write}$ (speed-up). 
This means a shorter pulse is favorable for producing current fluctuation at approximately $P_{\rm write} \approx 1/2$. 
{
The applied voltage to the MRAM cell is controlled by checking the `0' and `1' balance of
 the random data. Here, we prepared three typical types of data (Table II). Data $A$ (balanced data) is 
obtained after repeating the adjustment process of voltage control and the estimation of the `0' and `1' balance. 
Data $B$ (unbalanced data) is obtained by slightly changing the applied voltage of data $A$ in order to 
lose the balance. Data $C$ (slightly unbalanced data with a higher speed) is obtained by 
adjusting the applied voltage at a higher pulse generation rate.}

\begin{figure}
\centering
\includegraphics[width=8cm]{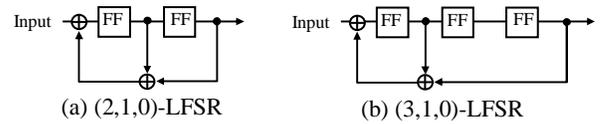}
\vspace*{2cm}
\caption{Small size LFSRs that improve original random signal. $N$ bit LFSR contains 
$N$ FFs. (a) 2-bit LFSR and (b) 3-bit LFSR~\cite{Schneier}.}
\label{LFSR}
\end{figure}

\begin{figure}
\centering
\includegraphics[width=5.5cm]{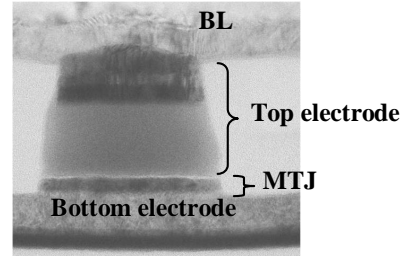}
\vspace*{4cm}
\caption{Cross-sectional TEM image of MTJ patterned by ion-milling method. 
The diameter of the element is 55 nm. MTJ contains the capping layer /
perpendicular reference layer /MgO / Fe-based L10-alloy /underlayer. }
\label{SEM}
\end{figure}

\begin{figure}
\centering
\includegraphics[width=6cm]{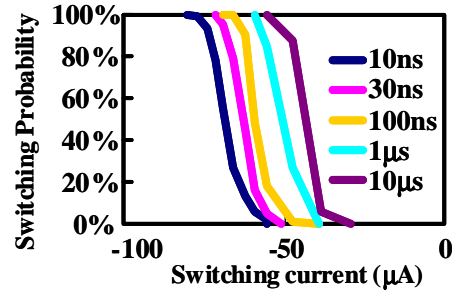}
\vspace*{4cm}
\caption{Switching probability as a function of switching current at various pulse durations $t_{\rm write}$.
The gentle tail part of the slope of shorter pulse such as $t_{\rm write}=10$~ns is considered to be favorable for generating 
random numbers, because it contributes to producing larger current fluctuations at approximately $P_{\rm write}\approx 1/2$.}
\label{pulse}
\end{figure}

\section{Results}
\begin{figure}
\centering
\includegraphics[width=8.5cm]{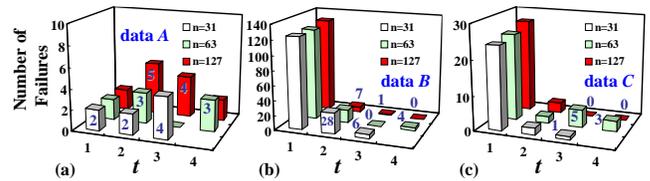}
\vspace*{2.5cm}
\caption{Result of statistical test (NIST800-22). Number of failures after 
ECC post processing for (a) good original random number bits (data $A$),
(b) unbalanced bits (data $B$) and (c) slightly unbalanced with higher speed (data $C$), 
as a function of error correcting capability $t$ for $n=31, 63$ and 127 BCH codes. 
Note that raw bit sequence just after MRAM unit fails in almost all these tests. That is, 
the number of failures for raw data is close to 160. }
\label{result1}
\end{figure}

\begin{figure}
\centering    
\includegraphics[width=7cm]{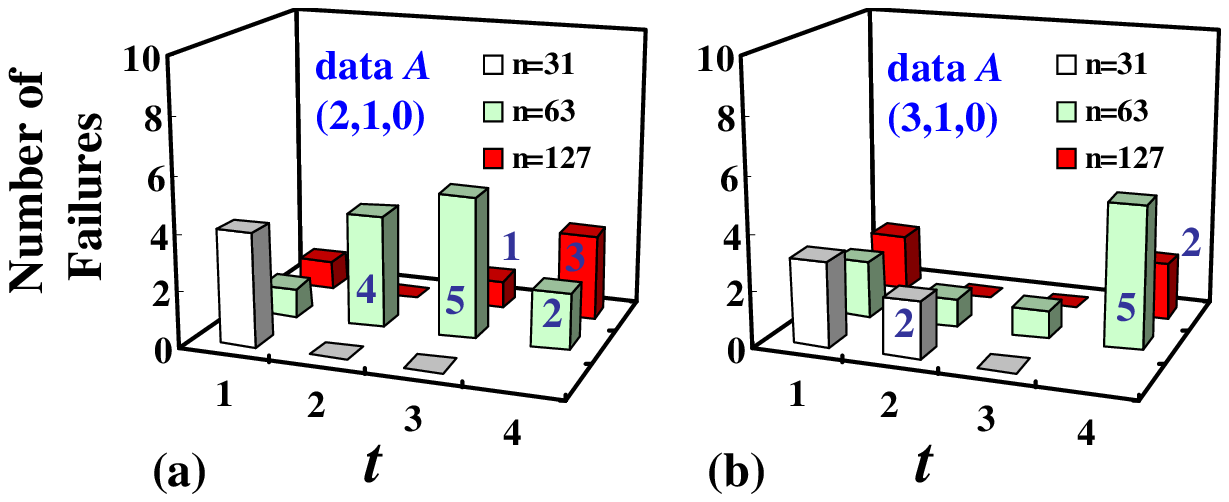}
\vspace*{3.1cm}
\caption{Result of statistical test (NIST 800-22) for data $A$ (balanced bits)
coupled with 2-bit LFSR (a) and 3-bit LFSR (b).
Both LFSRs improve the quality of randomness compared with Fig. \ref{result1} (a).}
\label{result2}
\end{figure}

\begin{figure}
\centering 
\includegraphics[width=7cm]{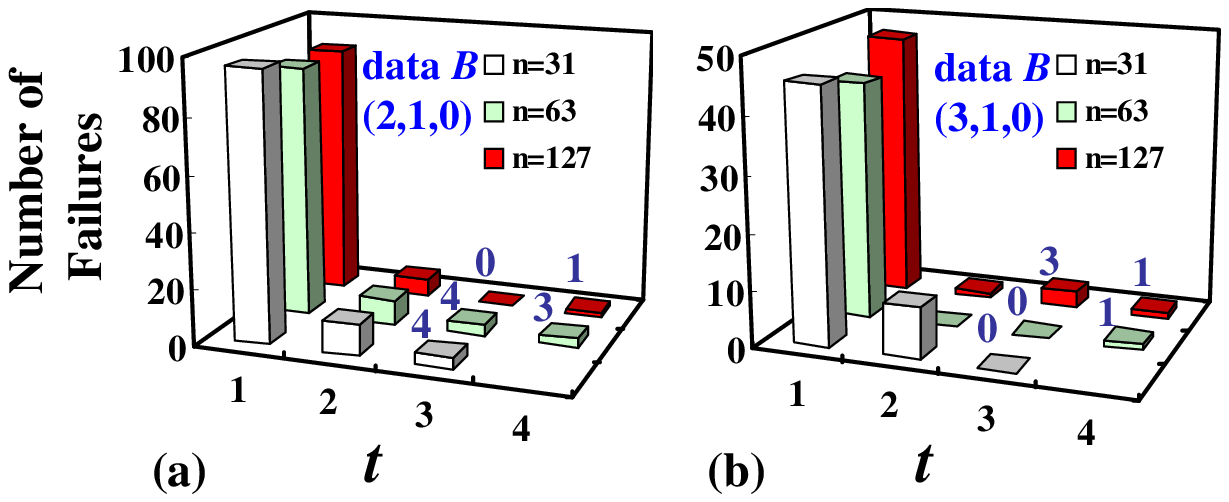}
\vspace*{3.1cm}
\caption{Result of statistical test (NIST 800-22) for data $B$ (unbalanced bits)
coupled with 2-bit LFSR (a) and 3-bit LFSR (b).
Both LFSRs improve the quality of randomness compared with Fig. \ref{result1} (b).}
\label{result3}
\end{figure}

\begin{figure}
\centering 
\includegraphics[width=7cm]{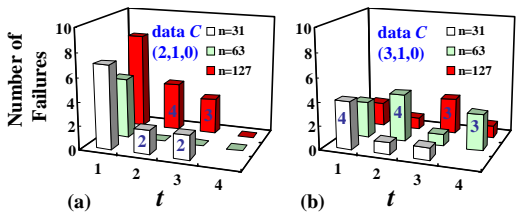}
\vspace*{3.1cm}
\caption{Result of statistical test (NIST 800-22) for data $C$ (slightly unbalanced bits with higher speed)
coupled with 2-bit LFSR (a) and 3-bit LFSR (b).
Both LFSRs improve the quality of randomness compared with Fig. \ref{result1} (c).}
\label{result4}
\end{figure}

Figure \ref{result1}-\ref{result4} show the results of statistical tests of randomness after ECC post processing for $n=31, 63$, and 
127 codes (Table~\ref{BCHcode}) as a function of the error-correcting capability $t$. 
For the three types of data (Table II), a rejection method compresses those data to 24.8 \% (data $A$), 20.0 \% (data $B$), and 23.1 \% (data $C$), respectively. 
In contrast to the rejection method, for ECC post processing, for example, for $t=2$,  
these rates are fixed to $k/n$ and given as 90.0 \% ($n=127$), 80.0 \% ($n=63$), 
and 67.7 \% ($n=31$). The statistical test suite NIST PUB SP 800-22~\cite{NIST} that contains 16 types of tests and judges 160 test 
results is used for a million bits set to a $P$-value of 0.01{~\cite{Pvalue}. }
Although there are several kinds of $Pass/Fail$ 
assignments for this test suite, we simply count the number of test failures and judge $Pass$ if it is 2 or less. 

Figure \ref{result1} shows results for PPC using ECC for balanced data (data $A$), unbalanced data with $t_{\rm write}$ =30~ns (data $B$), and 
slightly unbalanced with higher speed of $t_{\rm write}$ =10~ns (data $C$).
These results prove the improvement ability of the ECC 
post processing. In particular, unbalanced data (data $B$ and $C$) are greatly improved by this method.
Note that the raw bit sequence just after MRAM unit fails in almost all these tests. 
Moreover, we could not see any specific correlation between improvement and the error-correcting capability $t$. 
This does not follow the prediction of ref.~\cite{Lacharme}
in which better random numbers are obtained as $t$ (or $d$) increases ($d$ and $t$ has a relation $d>2t$). 
Because RNG speed decreases 
at rate $k/n$, and $n-k$ extra redundant memory cells are required, these results show that 
a small $t$ can be chosen. 

Figures \ref{result2}-\ref{result4} show results of ECC post processing with (2,1,0)- and (3,1,0)-LFSR.
We examined several lengths of LFSR and found that a length of two or three LFSRs is sufficient.  
That is, coupling with longer bit LFSRs such as  $(4,1,0)$, $(7,1,0)$, {\it etc}.
does not always show a greater improvement than 2-bit and 3-bit LFSRs.
This might be because LFSR itself does not increase the entropy of randomness and just mix random bit sequences.
If transistors can be set to adjust the generating polynomial [eq.(1)] depending on original random signals, 
the most appropriate ($n$,$k$,$t$) can be selected. 
When results of data $A$ are compared with those of data $C$,  it can be seen that 
the acceleration of current pulse from $t_{\rm write}$ =30~ns to $t_{\rm write}$=10~ns increases the quality 
of randomness by ECC post processing. This is considered to be related 
to the slope of switching probability (Fig. \ref{pulse}).
Finally, we estimate the generation speed of the present RNG.
When the read time is 10~ns and four clocks are required for outputting one-bit random number, 
25 MHz generation speed is expected for a RNG.
Table III shows that the estimated generation speeds of MRAM-based RNGs are much faster than those of 
state-of-the-art RNGs reported in ref.~\cite{Matsumoto} and \cite{Brederlow}.

\begin{table}[h]
\caption{Comparison of generation speeds of RNGs.}
\begin{tabular}{cc}
\hline
   RNG  & Speed (MHz) \\
\hline
Ref. \cite{Matsumoto} &  2 \\
\hline
Ref. \cite{Brederlow} & 0.2 \\
\hline
MRAM & 25 \\
\hline
\end{tabular}
\end{table}

{
\section{Discussion}
The reason why the PPC using ECC is effective for MRAM RNG is considered to be that 
MTJ directly outputs discrete two values of current. We also applied this method to 
RNG based on transistor noise~\cite{Matsumoto,Fujita}, in which the current noise signal is a continuous variable and 
should be converted to digital signals by an analog circuit. It is found that we need to apply 
this PPC a couple of times for RNG based on transistor noise to pass statistical tests. 
Thus, it seems that this method is less effective for RNG based on transistor noise.
Thus, direct observation of two distinct physical quantities in MTJ is considered to be related to the efficiency of the PCC using ECC.

We experimentally showed that small error-correcting capability is sufficient such as $t \le2$. As noted in the 
introduction, Lacharme only showed that ECC theoretically improved the balance between `0' and `1'\cite{Lacharme}. 
The theoretical explanation why a small $t$ is sufficient to pass the various statistical tests is a future problem.
 
Power consumption of widely used oscillator-based RNGs can be estimated to be more than 1 mW/bit. 
From the 2009 International Technology Roadmap for Semiconductors (ITRS)~\cite{itrs}, 
the write energy of MTJ is estimated as ${10}^{-13}$ J for one bit. 
If we take 2~ns as a write time, then the power consumption of MTJ is 0.05 mW/bit. 
Thus, the power consumption of MRAM RNGs is lower than that of currently used RNGs. 
In the near future, power consumption will be further reduced by decreasing switching current.

In general, random data are not very large in mobile applications and are used temporarily. 
Thus, the coexistence of memory data and random data is not a problem as long as each memory address is controlled.
}

\section{Conclusions}
We presented a novel MRAM-based RNG using ECC
in which ECC post processes random numbers 
in addition to fulfilling the conventional function of correcting memory bit errors.
We have proven that the MRAM-based RNG using ECC post processing efficiently 
increases both randomness quality and generation speed.

\section*{Acknowledgments}

The authors would like to thank A. Nishiyama, J. Koga,
S. Kawamura, K. Abe, S. Yasuda, K. Nomura, and H. Uchikawa for fruitful discussions.



\begin{thebibliography}{99}
\bibitem{Pvalue}
$P$-value is associated with a test statistic and is a probability that a perfect random number 
generator would have produced a sequence less random than the sequence that was tested, 
given the kind of nonrandomness assessed by the test~\cite{NIST}. If the $P$-value equals 1, 
the sequence appears to have perfect randomness and if the $P$-value equals 0 the 
sequence appears to be completely nonrandom.
\end{thebibliography}
\end{document}